\RequirePackage{lineno}
\documentclass[superscriptaddress,aps,preprint,amsmath,amssymb,floatfix]{revtex4}
\usepackage{graphicx,morefloats}
\usepackage{subfigure} 
\usepackage{color,soul}
\usepackage{bm}
\usepackage{amsmath}
\usepackage{amssymb}
\usepackage{times}
\usepackage{hyperref}
\usepackage{bbm}

\newcommand{\cepd}{Ce$_3$Bi$_4$Pd$_3$}
\newcommand{\cept}{Ce$_3$Bi$_4$Pt$_3$}
\newcommand{\lapd}{La$_3$Bi$_4$Pd$_3$}

\begin{document}

\hyphenation{va-ni-sh-ing to-po-lo-gy}

\begin{center}

\thispagestyle{empty}

{\Large\bf Controlling electronic topology in a strongly correlated\\  electron system}\\[0.3cm]

Sami Dzsaber$^1$, Diego A.\ Zocco$^1$, Alix McCollam$^2$, Franziska
Weickert$^3$, Ross McDonald$^3$, Mathieu Taupin$^1$, Xinlin Yan$^1$, Andrey
Prokofiev$^1$, Lucas M.\ K.\ Tang$^2$, Bryan Vlaar$^2$, Laurel E.\ Winter$^3$, Marcelo Jaime$^3$, Qimiao Si$^4$, and Silke Paschen$^{1,\ast}$\\[0.1cm]

$^1$Institute of Solid State Physics, Vienna University of Technology, 1040 Vienna, Austria\\
$^2$High Field Magnet Laboratory (HFML-EMFL), Radboud University,\\ 6525 ED Nijmegen, The Netherlands\\
$^3$Los Alamos National Laboratory, Los Alamos, NM 87545, USA\\
$^4$Department of Physics and Astronomy, Rice Center for Quantum Materials, Rice University, Houston, TX 77005, USA\\[-1.2cm]

\end{center}

\vspace{1cm}

\noindent $^{\ast}$Corresponding author. paschen@ifp.tuwien.ac.at

\vspace{1cm}

\noindent{\bf Combining strong electron correlations
\cite{Tok00.1,Loe07.1,Kei15.2,Pas21.1} and nontrivial electronic topology
\cite{Nar21.1} holds great promise for discovery. So far, this regime has been
rarely accessed and systematic studies are much needed to advance the field.
Here we demonstrate the control of topology in a heavy fermion system. We use
magnetic field to manipulate Weyl nodes in a Weyl-Kondo semimetal
\cite{Dzs17.1,Lai18.1,Dzs21.1}, up to the point where they annihilate in a
topological quantum phase transition. The suppression of the topological
characteristics occurs in an intact and only weakly varying  strongly correlated
``background''. Thus, topology is changing {\em per se} and not as a consequence
of a change of the correlation state, for instance across a magnetic, electronic
or structural phase transition. Our demonstration of genuine topology tuning in
a strongly correlated electron system sets the stage for establishing global
phase diagrams of topology, an approach that has proven highly valuable to
explore and understand topologically trivial strongly correlated electron
systems \cite{Tok00.1,Loe07.1,Kei15.2,Pas21.1}. Our work also lays the ground
for technological exploitations of controlled electronic topology.}
\newpage

Key to uncovering the richness of phenomena displayed by strongly correlated
electron systems and to understanding the underlying mechanisms has been
the great tunability of these systems, which enabled a systematic
exploration of the landscape of their low-temperature phases
\cite{Tok00.1,Loe07.1,Kei15.2,Pas21.1}. As an example, such studies lead to the
discovery that high-temperature superconductivity is in many, if not all cases
an emergent phase stabilized by quantum critical fluctuations
\cite{Mat98.1,Mic19.1}, and that critical electron delocalization transitions
\cite{Pas04.1,Pro20.1} may play an important role therein \cite{Bad16.1}. Such
deep insight is indispensable to tailoring properties at will, and ultimately
exploiting them for applications.

Electrons are strongly correlated if their mutual Coulomb repulsion reaches
or exceeds the same order as their kinetic energy. In bulk materials, this
regime is typically realized by means of $f$ and/or $d$ orbitals with a suitable
strength of orbital overlap. Extremely flat bands, renormalized by orders of
magnitude compared to simple metals and pinned to the Fermi energy can be
achieved via the Kondo effect, driven by a spin exchange interaction
between localized (typically $4f$) and itinerant ($s$, $p$, $d$) electrons
\cite{Hew97.1}. Recently, it has been demonstrated that such very strongly
correlated electron systems may also exhibit extreme signatures of nontrivial
topology \cite{Dzs21.1}. This raises the question as to whether the excellent
tunability in terms of correlation physics can also be exploited to control
topology {\em per se}. 

Our work shows that this can indeed be achieved. Upon tuning the Weyl-Kondo
semimetal Ce$_3$Bi$_4$Pd$_3$ \cite{Dzs17.1,Lai18.1,Dzs21.1} by magnetic field
we observe a continuous suppression of the giant topological response
associated with the material's Weyl nodes, and the annihilation of the nodes in
a topological quantum phase transition. This happens in a smoothly
varying correlated background. We expect this discovery to have broad
implications, both for the design and understanding of strongly correlated
topological materials and for technological exploitations.

\begin{figure*}[t]
\begin{center}
\vspace{-1cm}

\includegraphics[width=1.\textwidth]{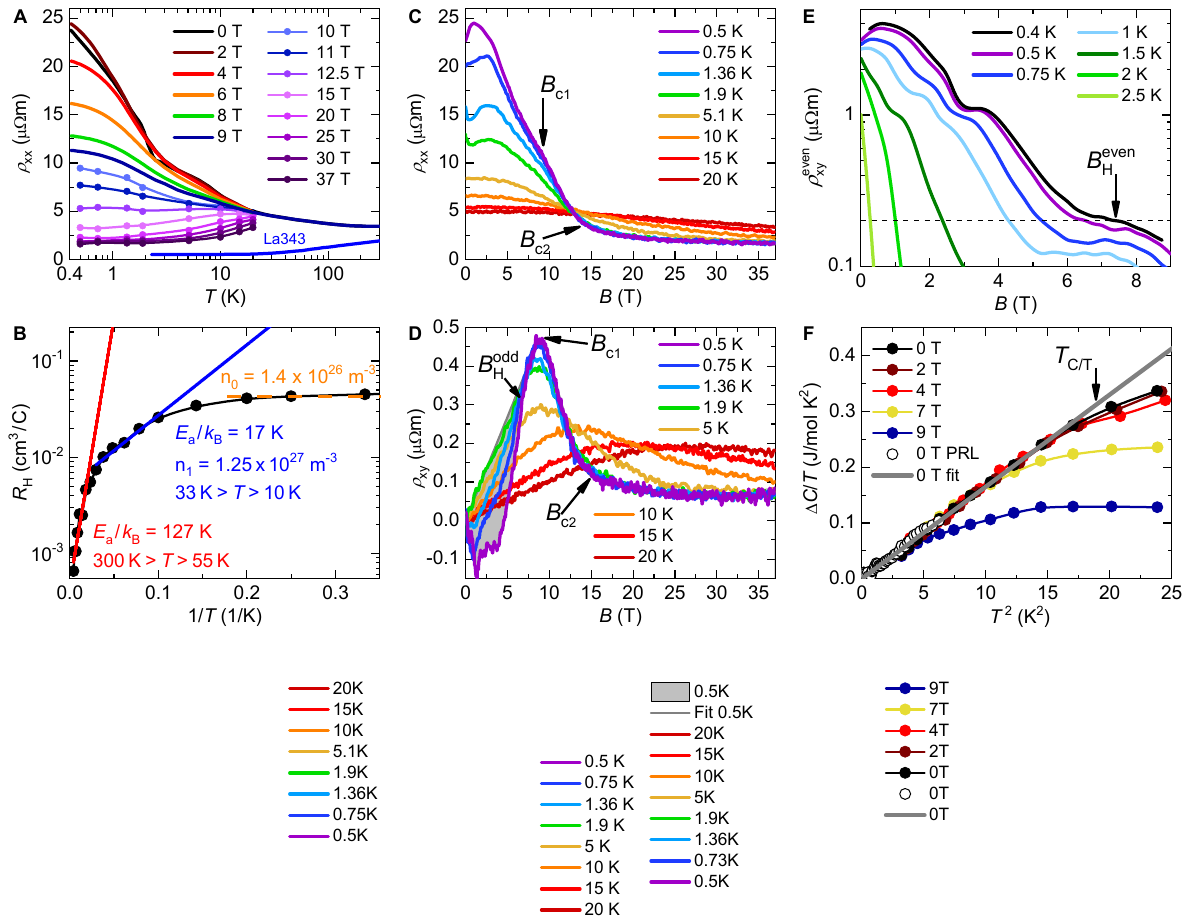}

\vspace{-0.4cm}

\caption[width=1\textwidth]{\label{Fig1} {\bf Data overview and
characteristic transport and thermodynamic scales of Ce$_3$Bi$_4$Pd$_3$.} ({\bf
A}) Temperature-dependent electrical resistivity at various fixed magnetic
fields. For $B>9$\,T, iso-$B$ cuts (dots) were taken from panel {\bf C}. The
zero-field resistivity of the nonmagnetic reference compound La$_3$Bi$_4$Pd$_3$
is shown for comparison. The small kink seen near 2.5\,K in the zero-field
data is associated with the onset of a giant spontaneous Hall voltage that, due
to the large associated Hall angle,  leaves an imprint also on the longitudinal
resistivity \cite{Dzs21.1}. The low-field data are taken from
Ref.\citenum{Dzs21.1}. ({\bf B}) Arrhenius plot of the linear response
normal (antisymmetrized) Hall coefficient $R_{\rm H}$ of Ce$_3$Bi$_4$Pd$_3$
(black symbols). The black line is a guide to the eyes. Red and blue lines
correspond to fits with $R_{\rm H} = R_{\rm H,i} \exp{[E_{\rm a}/(k_{\rm
B}T)]}$, where $n_i = 1/(R_{\rm H,i}e)$ is the charge carrier concentration in a
simple one-band model. Below 10\,K, the data saturate to a constant value
(dashed orange line), with the charge carrier concentration $n_0$ (again in a
one-band model). ({\bf C}) Magnetoresistance isotherms in fields up to $B =
37$\,T, for various temperatures between 0.5 and 20\,K. ({\bf D}) Normal
(antisymmetrized) Hall resistivity isotherms at the same fields and
temperatures. $B_{\rm c1}$ and $B_{\rm c2}$ are determined in Fig.\,\ref{Fig3}.
Below $B_{\rm H}^{\rm {odd}}$ (shown for the 0.5\,K data), the anomalous Hall contribution (grey shading)
leads to a deviation from the initial linear-in-$B$ normal Hall resistivity by more than 5$\%$. ({\bf E})
Even-in-field (symmetrized) Hall resistivity isotherms at low temperatures and
fields. Above $B_{\rm H}^{\rm {even}}$ (shown for the 0.4\,K data) this Weyl-node derived signal drops to
below 0.2, which is 5\% of the maximum of the lowest-temperature isotherm (data from Ref.\citenum{Dzs21.1}). ({\bf F})
Temperature-dependent electronic specific heat coefficient (see Supplementary
Sect.\,\ref{SM_CvsT} for details) vs $T^2$ at various fixed fields (open symbols
are from Ref.\citenum{Dzs17.1}). Above $T_{\rm C/T}$ (shown for the 0\,T data), the data deviate by more
than 5\% from the low-temperature $T^2$ fit, which represents the linear Weyl
dispersion. The spurious low-field features in the low-temperature resistivity isotherms (Fig.\,\ref{Fig1}C) are imprints of the large Hall contributions, caused by the current path redistribution due to the large Hall angle \cite{Dzs21.1}.}
\end{center}
\end{figure*}

Figure~\ref{Fig1} gives an overview of our electrical transport and specific
heat data. Salient transport features in zero magnetic field are an electrical
resistivity that increases moderately with decreasing temperature
(Fig.\,\ref{Fig1}A) and a linear-response Hall coefficient with two ranges of
thermally activated behavior and a saturation to a constant value at the lowest
temperatures (Fig.\,\ref{Fig1}B). A plausible interpretation, which will receive
further support from the field-dependent data presented below, is that this
behavior results from a (pseudo)gapped background density of states, within
which a narrow (Kondo insulator) gap forms at lower temperatures, against which
a small residual density of states associated with the Fermi-level-bound Weyl nodes becomes apparent at the lowest temperatures. The presence of a (pseudo)gap
in the noninteracting density of states is in agreement with density functional
theory (DFT) calculations, although the theoretical gap size is considerably
larger \cite{Tom18.1}.

\begin{figure*}[t]
\begin{center}
\includegraphics[width=1\textwidth]{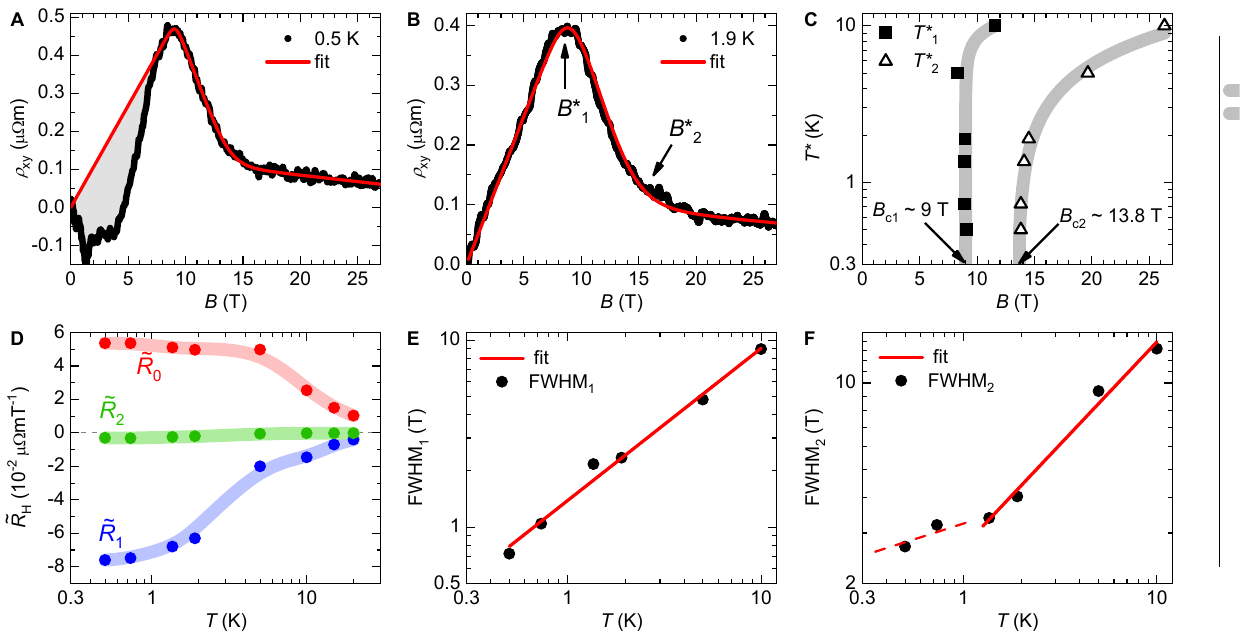}
\caption[width=1\textwidth]{\label{Fig3}{\bf Two-stage Fermi surface reconstruction in Ce$_3$Bi$_4$Pd$_3$.} ({\bf A}) Hall resistivity vs applied magnetic
field at 0.5\,K, together with the best fit according to the two-crossovers
model (see Supplementary Sect.\,\ref{SM_Hall}). The deviation of the data from the initial linear behavior
(shaded grey area) is due to a Berry curvature-derived anomalous Hall
contribution (Ref.\citenum{Dzs21.1}). ({\bf B}) Same as (A) at 1.9\,K. The
Berry-curvature derived contribution is essentially absent at this temperature.
The two characteristic fields $B^{\ast}_1$ and $B^{\ast}_2$ mark the positions
of the crossovers between the different regimes. ({\bf C}) The crossover fields
$B^{\ast}_1(T)$ and $B^{\ast}_2(T)$ determined from the $\rho_{xy}(B)$ fits are
plotted as $T^{\ast}_1(B)$ and $T^{\ast}_2(B)$, respectively, in a
temperature-magnetic field phase diagram. They extrapolate, in the
zero-temperature limit, to $B_{\rm{c1}}$ and $B_{\rm c2}$ and separate
three regimes of simple (linear-in-$B$) normal Hall resistivity. ({\bf D})
Differential Hall coefficients $\tilde{R}_0$, $\tilde{R}_1$, and $\tilde{R}_2$
of these three regimes, determined from the $\rho_{xy}(B)$ fits, as function of
temperature. The lines are guides to the eyes. ({\bf E}) Width of the first
crossover FWHM$_1$, representing the full width at half maximum of the second
derivative of the two-crossovers fit function. The straight line is a pure power
law fit, FWHM$_1 \propto T^p$, with $p=0.81$. It describes the data down to the
lowest temperature, evidencing that the carrier concentration changes abruptly
at $T=0$. ({\bf F}) Same as in (E) for the second crossover. The straight
full line is a pure power law fit, FWHM$_2 \propto T^p$, with $p=0.71$ to
the data above 1\,K. At lower temperatures, the rate of decrease is
somewhat reduced (red dashed line).}
\end{center}
\end{figure*}

The application of magnetic fields gradually suppresses the low-temperature
resistivity upturn. This is the case until, ultimately, metallic behavior is seen, albeit
with higher resistivity and a different temperature dependence than in the
nonmagnetic reference compound La$_3$Bi$_4$Pd$_3$ (Fig.\,\ref{Fig1}A). This indicates that Kondo physics is at play even at our largest field of 37\,T. Isothermal magnetic-field-dependent measurements of the electrical
resistivity $\rho_{xx}(B)$ (Fig.\,\ref{Fig1}C) and Hall resistivity
$\rho_{xy}(B)$ (Fig.\,\ref{Fig1}D) reveal that this field-induced transformation
occurs in two stages. The signatures thereof are most pronounced in the
lowest-temperature data. Here, the electrical resistivity displays a shoulder at
about 9\,T and a crossover to almost field-independent behavior at about 14\,T
(see arrows labeled $B_{\rm c1}$ and $B_{\rm c2}$, respectively, and Supplementary Sect.\,\ref{SM_Rxx_Bump} for further analyses). The corresponding
signatures in the Hall resistivity are kinks at the same two fields. A
quantitative analysis, presented further below (Fig.\,\ref{Fig3}), reveals that
this behavior reflects a two-stage Fermi surface reconstruction at two quantum phase transitions.

We first examine the effect the magnetic field has on the material's
characteristics of nontrivial topology. The most direct signature is a giant
spontaneous as well as even-in-field nonlinear Hall effect, which evidences
Berry curvature singularities from Weyl nodes in close vicinity to the Fermi
energy \cite{Dzs21.1}. Somewhat more indirect evidence is a
temperature-dependent electronic specific heat that varies linearly in $T^3$,
with a slope that even surpasses the (Debye-like) phonon contribution, and
evidences linearly-dispersing electronic bands with ultralow velocity
\cite{Dzs17.1,Lai18.1}. Together they have established the inversion
symmetry (IS)-broken (noncentrosymmetric and nonsymmorphic) but time reversal
symmetry (TRS)-preserving heavy fermion compound Ce$_3$Bi$_4$Pd$_3$ as a model
case of a strongly correlated topological semimetal
\cite{Dzs17.1,Dzs21.1}.

In Fig.\,\ref{Fig1}E we show how the Weyl-Kondo semimetal Hall signature changes
under magnetic field tuning. The even-in-field (symmetrized and corrected for
contact misalignment \cite{Dzs21.1}) isothermal Hall resistivities $\rho_{\rm
xy}^{\rm even}$ are successively suppressed by magnetic field. The apparent fine
structure in this suppression, seen in the isotherms below 2\,K, may reflect the
theoretically predicted rich sequence of Weyl node motion under magnetic
field tuning \cite{Gre20.2,Gre20.1x}, which is an interesting topic for further
investigations. Here, we focus on the ultimate total suppression of the effect,
which occurs at $B_{\rm H}^{\rm even}$ (see arrow in Fig.\,\ref{Fig1}E for the
lowest temperature isotherm). Also the linear-in-$T^3$ electronic specific heat
(or linear-in-$T^2$ electronic specific heat coefficient $C/T$) is successively
suppressed by magnetic fields, which we quantify by the parameter $T_{\rm C/T}$
(see arrow in Fig.\,\ref{Fig1}F on the zero-field curve). That no fine structure
is seen here is consistent with theoretical expectation because (linearly
dispersing) Weyl bands contribute to the $C/T \sim T^2$ signal regardless of
their position in momentum space, as long as the energy of the Weyl nodes
does not change. Finally, we point to another feature that accompanies these
two key signatures. It appears as an anomaly in the (normal, antisymmetrized and
thus odd-in-field) Hall resistivity isotherms (see arrow denoting $B_{\rm
H}^{\rm odd}$ as upper end of the grey shading, marked on the lowest-temperature
isotherm in Fig.\,\ref{Fig1}D) and is known as the anomalous topological Hall
effect in TRS-broken Weyl semimetals \cite{Nag10.1}. It is associated with a
magnetic field-induced even-in-momentum Berry curvature, which is in addition to
the intrinsic (zero-field) odd-in-momentum Berry curvature of Ce$_3$Bi$_4$Pd$_3$
(see also Ref.\citenum{Dzs21.1}).

All these results together establish that magnetic field quenches the topological response, likely via a process that moves the Weyl nodes at equal energy in momentum space. This raises the following questions: Which mechanism underlies this effect? Does magnetic field suppress the Kondo effect just as increasing the temperature above the Kondo coherence scale does \cite{Dzs21.1}, thereby removing the correlated electrons and thus the basis for the Weyl-Kondo semimetal formation? Or did we succeed to annihilate the Weyl nodes in an intact Kondo coherent system? The Kondo-like electrical resistivity in 37\,T suggests that some form of Kondo correlations persist. To show that the magnetic field indeed controls Kondo-driven Weyl nodal excitations, however, what needs to be established is that the Kondo effect as realized in zero magnetic field operates over the entire field range with topological response. We now turn to the search for such evidence.

We start with a quantitative analysis of the
Hall resistivity isotherms of Fig.\,\ref{Fig1}D. As established previously
\cite{Pas04.1}, when magnetic field drives transitions between ground states
with different Fermi volumes, the resulting (finite temperature) crossovers
manifest as (broadened) kinks in the (normal) Hall resistivity $\rho_{xy}(B)$, and
(broadened) steps in the differential Hall coefficient $\tilde{R}_{\rm H}(B) =
\partial\rho_{xy}(B)/\partial B$. Such behavior has been observed in a number of
heavy fermion metals \cite{Pas04.1,Fri10.2,Cus12.1,Mar19.1} driven by magnetic
field across quantum critical points. From fits with a 
phenomenological crossover function \cite{Pas04.1} (see Supplementary
Sect.\,\ref{SM_Hall} for details) one can extract not only the
$\tilde{R}_{\rm H}$ values associated with the different phases, but also the
crossover fields $B^{\ast}$ and sharpnesses, quantified by the full width at
half maximum (FWHM). In Fig.\,\ref{Fig3}A,B we show two representative results of such fits, for data
at 0.5\,K (for which we
have subtracted the above-discussed anomalous topological Hall effect contribution) and 1.9\,K, respectively. Fits
of similar quality are obtained at all temperatures up to 10\,K
(Fig.\,\ref{Fig_SM_Hall}). At higher temperatures, we lose track of the
two-stage nature and thus this model does no longer give meaningful results. Note that anomalous (nontopological) Hall contributions and multiband
effects do not play significant roles here (see Supplementary Sects.\,\ref{SM_SkewScattering} and
\ref{SM_TwoBand}). The
temperature-dependent fit parameters are shown in Fig.\,\ref{Fig3}C-F. The two
crossover fields $B^{\ast}_1$ and $B^{\ast}_2$ (Fig.\,\ref{Fig3}B), determined
for all available isotherms, are plotted as characteristic temperatures
$T^{\ast}_1(B)$ and $T^{\ast}_2(B)$ in Fig.\,\ref{Fig3}C. The
extrapolations to $T=0$ of these curves identify $B_{\rm c1}$ and $B_{\rm c2}$
(see arrows in Fig.\,\ref{Fig3}C). Both crossovers sharpen considerably with
decreasing temperature (Fig.\,\ref{Fig3}E,F), indicating that the phase diagram
of magnetic field-tuned Ce$_3$Bi$_4$Pd$_3$ comprises three phases with distinct
Fermi volumes: a phase below $B_{\rm{c1}}$ with a small hole-like Fermi volume,
an intermediate-field phase between $B_{\rm{c1}}$ and $B_{\rm{c2}}$ with an even
smaller electron-like Fermi volume, and a high-field phase beyond $B_{\rm{c2}}$
with a much larger Fermi volume.

To elucidate the character of these phases further we have carried out torque
magnetometry measurements. At low magnetic fields, and in particular across $B^{\ast}_1(T)$, no sizable torque signal
is detected (Fig.\,\ref{Fig2}A), thus ruling out that a magnetic phase transition occurs at this field (see
Supplementary Sect.\,\ref{SM_Torque} and Fig.\,\ref{Fig_SM_torque}). A pronounced
torque signal appears only above about 14\,T; it corresponds to the onset of
nonlinearity in the magnetization \cite{Kus19.1}. Similar behavior, albeit with
a much larger magnetic field scale, is also seen in the canonical Kondo
insulator Ce$_3$Bi$_4$Pt$_3$, which we have studied for comparison
(Fig.\,\ref{Fig2}B). The corresponding characteristics in the first and second derivative with respect to
the magnetic field are a step-like increase and a maximum, respectively. For Ce$_3$Bi$_4$Pt$_3$, the step in the lowest temperature isotherm occurs at 38.9\,T (blue arrow in Fig.\,\ref{Fig2}C), which is close to the field where a Kondo insulator to metal transition has previously been
evidenced by a jump of the Sommerfeld coefficient \cite{Jai00.1}. The very similar feature seen for Ce$_3$Bi$_4$Pd$_3$ (red line in Fig.\,\ref{Fig2}C) then suggests that also in this system a Kondo insulator to metal transition takes place, albeit at much lower fields. As characteristic field $B_{\tau}$ of this transition, which is well-defined for all isotherms, we use the middle field at half height of the second derivative curves (Fig.\,\ref{Fig2}D).

We also performed temperature-dependent electrical resistivity measurements at
fields around this Kondo insulator to metal transition. On the high-field side
of the transition, we observe Fermi liquid behavior, $\rho = \rho_0 + AT^2$ (see
Fig.\,\ref{Fig2}E, bottom and left axes, for data at 15\,T). This confirms that
Ce$_3$Bi$_4$Pd$_3$ has indeed metallized. The $A$ coefficient measures the
strength of electronic correlations. Values in the range of several $\mu\Omega$cm/K$^2$,
as observed here, are typical of heavy fermion metals \cite{Kad86.1}. Thus, the
quenching of the Kondo insulator gap by the magnetic field has indeed (as already indicated by the high-field curves in Fig.\,\ref{Fig1}A) still not
suppressed the Kondo interaction. Thus, what happens at $B_{\rm c2}$ is a field-induced
Kondo insulator to heavy fermion metal transition. In Fig.\,\ref{Fig2}F we plot
the $A$ coefficient determined also for other fields (see
Fig.\,\ref{Fig_SM_rhoFL}) as function of magnetic field. Upon approaching the
transition from the high-field side, the $A$ coefficient increases, and is well
described by a divergence, $A \propto 1/(B-B_{\rm c2})$, with the critical
field $B_{\rm c2}=13.8$\,T from the Hall resistivity analysis (see caption of
Fig.\,\ref{Fig2} for details). Such behavior is known from heavy fermion metals
tuned by a magnetic field to a quantum critical point (QCP) \cite{Geg02.1,Mar19.1}. In this case, also
non-Fermi liquid (NFL) behavior should develop \cite{Si01.1,Cus03.1}, which we indeed
observe in the form of a linear-in-temperature resistivity at 15\,T, at
temperatures above the Fermi liquid behavior seen at the lowest temperatures
(Fig.\,\ref{Fig2}E, top and right axes). This confirms that, at 15\,T,
Ce$_3$Bi$_4$Pd$_3$ is slightly away from a QCP. Closer to the expected quantum
critical field $B_{\rm{c2}}$, the resistivity appears to be  influenced by the
nearby Kondo insulator phase, as seen by a rapid suppression of the $A$
coefficient and a crossover to $T^2$ behavior with a negative slope
(Fig.\,\ref{Fig_SM_rhoFL}H), as well as an increase of the residual resistivity
$\rho_0$ not only towards but even across $B_{\rm{c2}}$
(Fig.\,\ref{Fig_SM_rhoFL}I).  Whether these (nonmetallic) characteristics are generic to field-induced
Kondo insulator to heavy fermion metal transitions is an interesting topic for
future studies (we note that in a related pressure-induced transition in SmB$_6$ no such effects were seen \cite{Gab03.1,Zho17.2}, but this may reflect the need of extending those measurements down to the temperature range of dilution refrigerators). Clearly, they do not represent normal (metallic)
behavior and thus we define the upper boundary of Fermi liquid behavior
$T_{\rm{FL}}$ (see arrows in Figs.\,\ref{Fig2}E and \ref{Fig_SM_rhoFL}) only in
the field range where these effects have minor influence. We conclude that a heavy Fermi liquid phase exists at fields above
$B_{\rm{c2}}$. In turn, this implies that the Kondo effect as operating at $B=0$ remains intact across $B_{c1}$.

We are now in the position to construct a temperature-magnetic field phase
diagram that assembles the above discussed characteristics
(Fig.\,\ref{Fig4}A). At low fields, these are $T_{\rm C/T}(B)$, $T_{\rm H}^{\rm
even}(B)$, and $T_{\rm H}^{\rm odd}(B)$, all denoting
temperatures up to which, at a given field, a certain signature of Weyl-Kondo
semimetal behavior is detected. These scales all collapse at a critical field
$B_{\rm c1}$ of 9\,T. Importantly, the correlated background only gradually evolves across this field and no magnetic phase transition takes place. At high fields, we plot the scales $T_{\tau}(B)$ and
$T_{\rm FL}(B)$ associated with the torque anomaly and Fermi liquid behavior,
respectively, and also include the inverse of the Fermi liquid resistivity coefficient, $1/A$, which
hits zero at the effective mass divergence. In addition, we show the
$T^{\ast}_1(B)$ and $T^{\ast}_2(B)$ scales extracted from our quantitative
Hall resistivity analysis. Note that all these temperature scales are
crossovers; they do not represent phase boundaries in the thermodynamic sense as
none of the phases is associated with an order parameter.

In Fig.\,\ref{Fig4}B we plot the charge carrier concentration $n$ per Ce,
extracted in a single-band model from the lowest-temperature values of the
differential Hall coefficients (Fig.\,\ref{Fig3}D). It changes from about 0.01
holes below $B_{\rm c1}$, via 0.007 electrons between $B_{\rm c1}$ and $B_{\rm
c2}$, to 0.18 electrons beyond $B_{\rm c2}$. The 25-fold increase across $B_{\rm c2}$ is independent evidence for the above discussed Kondo insulator to heavy fermion metal transition. The sign change of $n$ and the more modest reduction of the absolute value of the carrier concentration across $B_{\rm c1}$ indicate a Fermi surface reconstruction that involves only a fraction of momentum space. The annihilation and associated gapping-out of Weyl nodes is exactly such a phenomenon. The field-independent charge carrier concentration at fields below $B_{\rm c1}$, evidenced by the simple linear-in-field behavior of the Hall resistivity in this regime (directly seen in Fig.\,\ref{Fig3}B), supports this assignment. This is because, theoretically, Weyl nodes in a Weyl-Kondo semimetal are expected to be moved in $k$ space under the action of Zeeman coupling associated with the magnetic field, but to remain at constant energy \cite{Gre20.2,Gre20.1x}, thereby also preserving the charge carrier concentration. Because a pair of Weyl and anti-Weyl nodes acts as particle and antiparticle, the only process that can remove them from the material is their mutual annihilation. This is what we propose to happen at $B_{\rm c1}$. Importantly, this takes place in a background of unchanged symmetry and with the Kondo effect continuing to operate. In other words, we have realized a controlled suppression of the Kondo-driven topological semimetal. A corollary is that we have isolated the tuning of topology from the tuning of correlation physics as such.

Our results are summarized in the schematic zero-temperature phase diagram in
Fig.\,\ref{Fig4}C. The low-field phase is a Weyl-Kondo semimetal (WKSM)
\cite{Dzs17.1,Lai18.1,Dzs21.1} that, as evidenced here, consists of Weyl nodes
situated within the narrow energy gap of a Kondo insulator. As function of
magnetic field a sequence of two Fermi volume-changing quantum phase transitions
is observed. At $B_{\rm c1}$, all signatures of the Weyl-Kondo semimetal
disappear as the Weyl nodes annihilate in a
topological quantum phase
transition. At $B_{\rm c2}$, the Kondo insulator with gapped-out Weyl nodes situated within its gap (denoted by the symbol
$\Delta$) transforms into a
heavy fermion metal (HF metal), in a transition that displays signatures of
quantum criticality and must thus be at least nearly continuous.

In summary, we have demonstrated the genuine control of Weyl nodes, with
unprecedented clarity and ease. The clarity is attributed to the fact that
in Ce$_3$Bi$_4$Pd$_3$ the Weyl nodes form within a (strongly correlated, Kondo
insulating) gapped state and are positioned in close vicinity of the Fermi
energy. Because there are no topologically trivial states in the gap, there
is no need to disentangle Weyl fermions from topologically trivial carriers,
which hampers the field of weakly interacting Weyl semimetals \cite{Arm18.1}.
The ease of control---namely that a rather modest field of 9\,T was not
only enough to manipulate the positions of the Weyl nodes in momentum space, but
even drive their annihilation---is a striking new result. It shows that the
well-known excellent tunability of (topologically trivial) strongly correlated
electron systems \cite{Tok00.1,Loe07.1,Kei15.2,Pas21.1} holds also
for topological features in such systems. In weakly interacting Weyl semimetals, by contrast, the tuning
by the Zeeman coupling appears to be negligible \cite{Ram18.1}; if changes of topology do occur, they are tied to changes in the material's broken symmetry state \cite{Sch18.1}. Similar contrast
between strongly correlated and weakly correlated topology also applies for
other control parameters. For instance, in sequences considered to be driven by
increasing spin-orbit coupling strength, a Weyl-Kondo semimetal to Kondo
insulator transition was observed in Ce$_3$Bi$_4$(Pd$_{1-x}$Pt$_x$)$_3$ with
increasing $x$ \cite{Dzs17.1}, whereas in the series of weakly interacting compounds
NbP -- TaP -- TaAs, ARPES could only detect a relatively modest change in Weyl
node separation (by 0.07 \AA$^{-1}$ across the whole series) \cite{Liu16.1}. The ability to continuously vary the magnetic field and thus the Zeeman coupling, as opposed to discrete steps in such substitution studies, was crucial for our investigation and makes this technique particularly powerful.

Our study lays the ground for establishing a global phase diagram---which has
been vital for understanding topologically trivial heavy fermion metals
\cite{Pas21.1}---for strongly correlated topological materials. Key open
questions to address are whether the phases and transitions discovered here
exist also in other materials and are thus universal, whether other exotic
phases, possibly without analogs in weakly interacting systems, can be found,
and whether quantum criticality plays an important role in stabilizing any of
these phases. The latter is hinted at by recent inelastic neutron scattering
experiments \cite{Fuh20.1x}.

Our findings may also guide investigations in related materials classes.
Much effort is currently devoted to artificial materials such as twisted bilayer
systems where correlations can be enhanced via a moir\'e potential
\cite{Bis11.1,Cao18.1,Cho21.1,Ken20.1x}. Additional tuning knobs in such
heterostructures are the dielectric displacement and electrostatic doping, which
may become powerful if further advances towards highly reproducible
structures can be accomplished. Finally, we point to the potential of
strongly correlated bulk materials such as Ce$_3$Bi$_4$Pd$_3$ for quantum
devices, where the robust and giant topological response---together with
the high level of topology control demonstrated here---opens entirely new
opportunities.

\newpage



%

\vspace{0.5cm}

{\noindent\large\bf Acknowledgements}\\
The authors wish to thank H.-H.\ Lai and S.\ Grefe for fruitful discussions and
G.\ Eguchi for support with the two-band analysis. We acknowledge the support of
the HFML, member of the European Magnetic Field Laboratory (EMFL). A portion of
this work was performed at the National High Magnetic Field Laboratory, which is
supported by the National Science Foundation Cooperative Agreement No.
DMR-1157490 and DMR-1644779, the State of Florida and the United States
Department of Energy. The work in Vienna was supported by the Austrian Science
Fund (DK W1243, I2535, I4047, and P29296). The work at Rice was in part
supported by the NSF (DMR-1920740) and the Robert A. Welch Foundation (C-1411),
the ARO (W911NF-14-1-0525).

\newpage

{\noindent\large\bf Author Contributions}\\
S.P.\ designed and guided the research. X.Y.\ and A.P.\ synthesized and
characterized the material. S.D., D.Z., A.M., F.W., R.M., L.T., B.V., L.E.W., and
M.J.\ performed the high-field experiments, M.T., S.D.\ and D.Z.\ the dilution
refrigerator experiments. S.D.\ analyzed the data, with contributions from D.Z.,
M.T., and S.P. The manuscript was written by S.P., with contributions from S.D., D.Z.\ and Q.S. All authors contributed to the discussion.
\vspace{0.5cm}

\newpage



\clearpage

\newpage


\begin{figure*}[t!]
\begin{center}
\includegraphics[width=1.\textwidth]{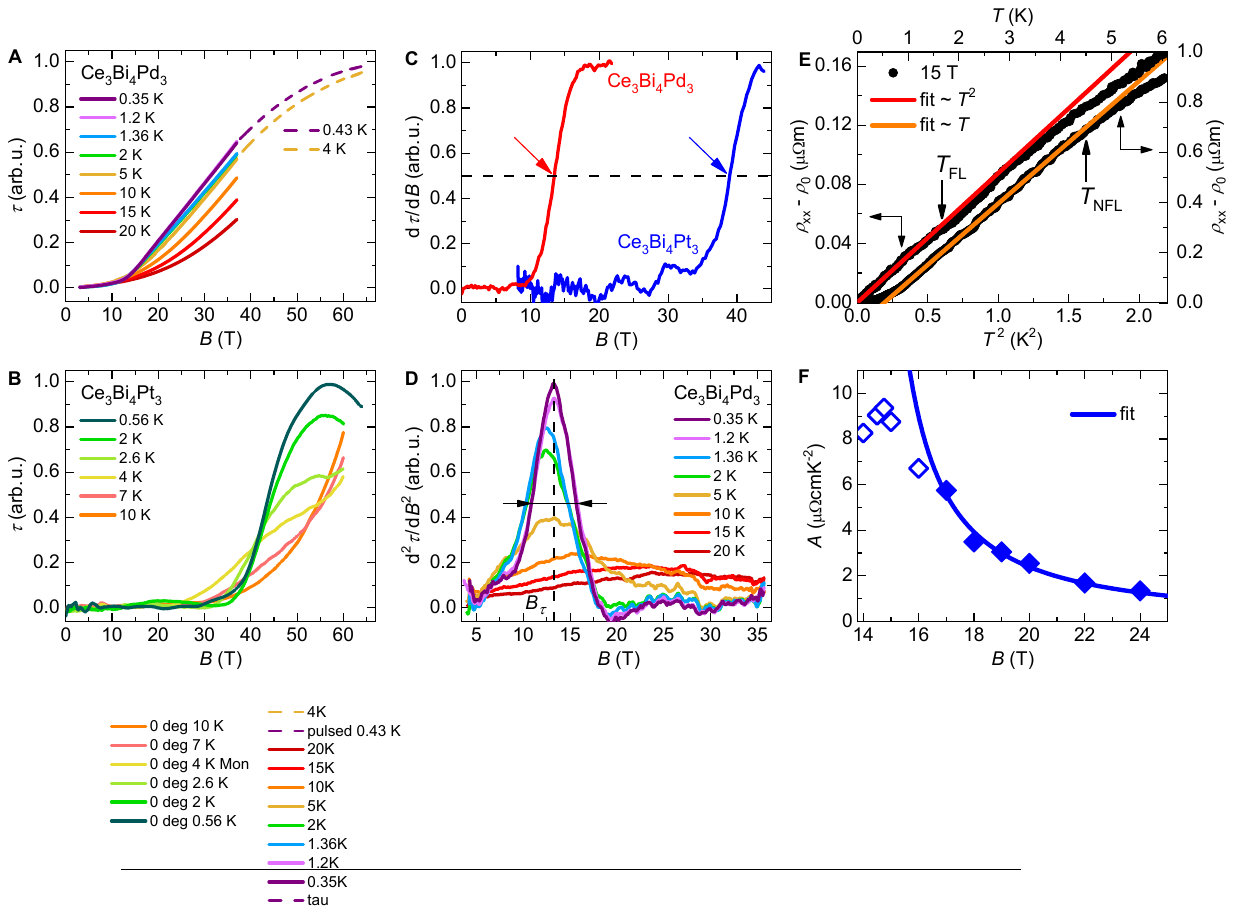}

\caption[width=1\textwidth]{\label{Fig2}{\bf Field-tuned Kondo insulator to
heavy fermion metal transition in Ce$_3$Bi$_4$Pd$_3$.} ({\bf A}) Magnetic
torque isotherms of Ce$_3$Bi$_4$Pd$_3$ at various temperatures as function of
applied magnetic field up to 37\,T and, for low temperatures, up to 65\,T. ({\bf
B}) Same as in (A) for the Kondo insulator Ce$_3$Bi$_4$Pt$_3$. ({\bf C}) Field
derivative of the magnetic torque isotherms of Ce$_3$Bi$_4$Pd$_3$ and
Ce$_3$Bi$_4$Pt$_3$ at the lowest temperatures, revealing similar
characteristics, albeit at different fields (the fields where the step-like
increases reach half height are indicated by arrows). ({\bf D}) Second field
derivative of the torque signal of Ce$_3$Bi$_4$Pd$_3$. We define the
characteristic field $B_{\tau}$ of the torque signal as the center of the width
at half maximum, as indicated by the dashed vertical line and the horizontal
arrows for the 0.35\,K isotherm. ({\bf E}) Electrical resistivity of
Ce$_3$Bi$_4$Pd$_3$ at 15\,T, displaying linear-in-$T^2$ Fermi liquid behavior
from the lowest temperature of 0.1\,K up to $T_{\rm FL}$ (bottom and left axes,
red line; above $T_{\rm{FL}}$, the data deviate by more than 5\% from the Fermi liquid law $\rho = \rho_0 + A T^2$) and linear-in-$T$ non-Fermi liquid behavior from somewhat above
$T_{\rm FL}$ up to $T_{\rm NFL}$ (top and right axes, orange line; see
Supplementary Sect.\,\ref{SM_FL} for details). ({\bf F}) Fermi liquid $A$
coefficient vs applied magnetic field. The data above 16\,T are well described
by $A \propto (B-B_{\rm c2})^{-p}$ with $B_{\rm c2}=13.8$\,T and $p \approx 1$.
Deviations at lower fields are attributed to proximity to the Kondo
insulating phase (see Supplementary Sect.\,\ref{SM_FL} and
Fig.\,\ref{Fig_SM_rhoFL}).}
\end{center}
\end{figure*}

\clearpage

\newpage


\begin{figure*}[t]
\begin{center}
\includegraphics[width=0.5\textwidth]{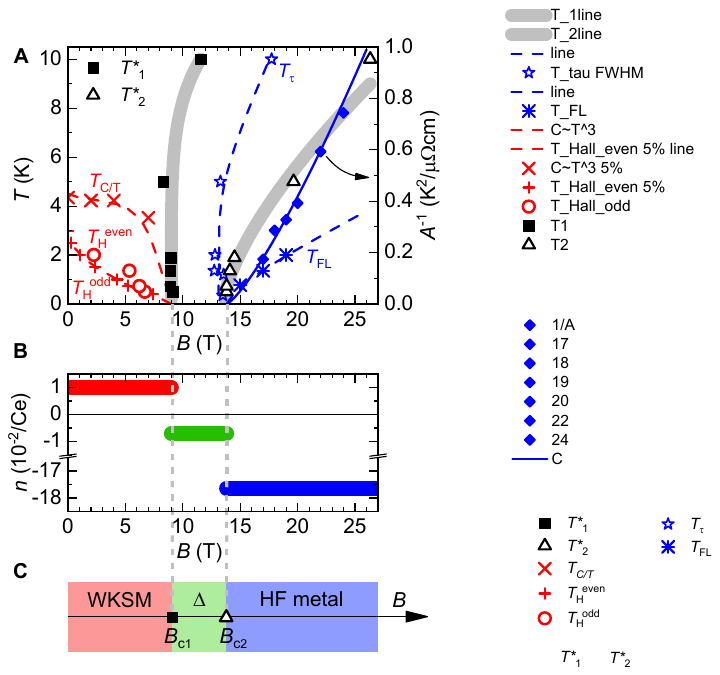}
\caption[width=1\textwidth]{\label{Fig4}{\bf Magnetic-field tuned phase diagram of Ce$_3$Bi$_4$Pd$_3$.} ({\bf A}) Temperature-magnetic field phase diagram with the crossover temperatures $T^{\ast}_1$ and $T^{\ast}_2$ (from Fig.\,\ref{Fig3}C), the temperature scales $T_{\rm C/T}$, $T_{\rm H}^{\rm {even}}$, and $T_{\rm H}^{\rm {odd}}$ (all from Fig.\,\ref{Fig1}) associated with the Weyl-Kondo semimetal phase, the characteristic scales $T_{\tau}$ of the torque signal and $T_{\rm{FL}}$ of Fermi liquid behavior (both from Fig.\,\ref{Fig2}) (all left axis). The inverse $A$ coefficient (from Fig.\,\ref{Fig2}) is plotted on the right axis, indicating an effective mass divergence at $B_{\rm c2}$. ({\bf B}) Effective charge carrier concentration in a single-band model, $n=1/(\tilde{R} e)$, at the lowest temperature of 0.5\,K (from Fig.\,\ref{Fig3}D), in the three different magnetic field ranges. ({\bf C}) Axis across a theoretical zero-temperature phase diagram. A topological quantum phase transition between a Weyl-Kondo semimetal (WKSM) and a phase with annihilated and thus gapped-out Weyl nodes ($\Delta$) occurs at $B_{\rm c1}$. The underlying Kondo insulator gap persists across both the WKSM and the $\Delta$ phase, and is quenched only at a quantum critical point at $B_{\rm c2}$ to reach a heavy fermion (HF) metal phase at high fields.}
\end{center}
\end{figure*}

\clearpage

\newpage



\newpage

{\noindent\large\bf Methods}

\noindent{\bf A. Synthesis} 

\noindent Single crystals of \cepd\,, \cept\, and of the nonmagnetic reference compound \lapd\ were grown using a modified Bi flux method, as described in Ref.\citenum{Dzs17.1}. The chemical composition and crystal structure of the samples were determined by energy dispersive x-ray spectroscopy and powder x-ray diffraction. Laue diffraction was utilized to determine the crystallographic orientation of selected samples.

\noindent{\bf B. Measurement setups}\\
\noindent{\bf High-field experiments}
\noindent Magnetoresistance, Hall effect, and torque magnetization measurements in DC fields up to 37\,T were performed at the HFML-EMFL facility at Nijmegen. Magnetotransport data was measured using Stanford Research SR830 lock-in amplifiers, with the measured voltage signal pre-amplified 100 times using Princeton Applied Research low-noise transformers. Electrical contacts where made by either spot welding or gluing with silver paint 12\,$\mu$m diameter gold wires to the samples in a 5-wire configuration. All displayed Hall resistivity curves were obtained by the standard antisymmetrizing procedure of the resistivity $\rho_{xy}^{\rm{meas}}$ measured across the Hall contacts, i.e., $\rho_{xy}(B) = [\rho_{xy}^{\rm{meas}}(+B) - \rho_{xy}^{\rm{meas}}(-B)]/2$. This cancels out both the spontaneous Hall effect (zero-field signal) and any even-in-field component. Torque and magnetization measurements in pulsed fields up to 65\,T were
performed at the NHMFL-LANL facility at Los Alamos. Magnetization data were
obtained only for \cept, using an extraction magnetometer and the
``sample-in/sample-out'' technique to separate the sample signal from the
background. In this case, the magnetic field was not aligned to any particular
crystallographic axis. In all torque experiments, piezo-cantilevers were used for enhanced sensitivity. Samples were attached to the levers with Dow Corning high vacuum grease. The measured torque signal was obtained after balancing a Wheatstone bridge containing the resistance of the sample's cantilever and a ``dummy'' (empty lever) resistor.  Due to the small samples required for this technique, the magnetic field was initially not aligned to any particular crystallographic axis. A sample rotator was used to scan the torque signal across different orientations. In all cases, temperatures down to 0.35\,K were obtained using a $^3$He cryostat.

\noindent{\bf Low-temperature experiments}
Additional magnetoresistance, Hall effect, and specific heat measurements were obtained at TU Vienna using a Quantum Design Physical Property Measurement System, equipped with $^3$He options. Low-temperature electrical resistivity measurements down to 70\,mK and in magnetic fields up to 15\,T were performed in an Oxford dilution refrigerator. 

\newpage

\pagestyle{empty}
\nolinenumbers



\addtocounter{figure}{-4} 
\addtocounter{equation}{0} 
\makeatletter
\renewcommand{\thefigure}{S\@arabic\c@figure}
\renewcommand{\theequation}{S\@arabic\c@equation}
\renewcommand{\thetable}{S\@arabic\c@table}

\noindent{\LARGE Supplementary Information}
\vspace{0.3cm}

\noindent The supplementary information contains the Sections
\vspace{-0.8cm}

\section{Evidence for a two-stage transition from electrical resistivity}\label{SM_Rxx_Bump}
\vspace{-0.6cm}

\section{Analysis of Hall resistivity data}\label{SM_Hall}
\vspace{-0.6cm}

\section{Anomalous Hall effect}\label{SM_SkewScattering}
\vspace{-0.6cm}

\section{Hall effect from multiple bands}\label{SM_TwoBand}
\vspace{-0.6cm}

\section{Analysis of torque magnetometry data}\label{SM_Torque}
\vspace{-0.6cm}

\section{Specific heat}\label{SM_CvsT}
\vspace{-0.6cm}

\section{Fermi and non-Fermi liquid behavior}\label{SM_FL}
\vspace{-0.4cm}

\noindent the Figures
\vspace{-0.5cm}

\begin{figure}[h!]
\caption{\label{Fig_SM_RhoKink} Field-dependent electrical resistivity.\hspace{14cm}}
\end{figure}
\vspace{-1cm}

\begin{figure}[h!]
\caption{\label{Fig_SM_RhoSlope} Temperature dependent iso-field electrical resistivity curves. \hspace{14cm}}
\end{figure}
\vspace{-1cm}

\begin{figure}[h!]
\caption{\label{Fig_SM_Hall} Two-stage crossover fits to the Hall resistivity. \hspace{14cm}}
\end{figure}
\vspace{-1cm}

\begin{figure}[h!]
\caption{\label{Fig_SM_Skew1} Estimation of skew scattering contribution to Hall effect from magnetization. \hspace{14cm}}
\end{figure}
\vspace{-1cm}

\begin{figure}[h!]
\caption{\label{Fig_SM_Skew2} Estimation of skew scattering contribution to Hall effect from magnetization and electrical resistivity. \hspace{14cm}}
\end{figure}
\vspace{-1cm}

\begin{figure}[h!]
\caption{\label{Fig_SM_2band} Estimation of contribution to Hall effect from multiband effects. \hspace{14cm}}
\end{figure}
\vspace{-1cm}

\begin{figure}[h!]
\caption{\label{Fig_SM_torque} Torque magnetometry of \cepd\ and \cept. \hspace{14cm}}
\end{figure}
\vspace{-1cm}

\begin{figure}[h!]
\caption{\label{Fig_SM_rhoFL} Analysis of Fermi liquid behavior in the electrical resistivity of \cepd. \hspace{14cm}}
\end{figure}
\vspace{-0.4cm}

\noindent and the Table
\vspace{-0.5cm}

\begin{table}[!h]
\caption{\label{SMtable1} Fit parameters for two-band analysis. \hspace{14cm}}
\end{table}

\end{document}